\documentclass[twocolumn]{revtex4-1}
\usepackage{graphicx}
\usepackage{xcolor}

\begin{document}

\title{Autonomous Brownian gyrators: a study on gyrating characteristics}

%\homepage[]{Your web page}
%\thanks{}
%\altaffiliation{}

\author{Hsin Chang}
\author{Chi-Lun Lee}
\email[]{lee.chilun@gmail.com}
\author{Pik-Yin Lai}
\author{Yung-Fu Chen}
\affiliation{Department of Physics, National Central University, Zhongli 32001, Taiwan}

\date{\today}

\begin{abstract}
We study the nonequilibrium steady-state (NESS) dynamics of two-dimensional Brownian gyrators under harmonic and nonharmonic potentials via computer simulations and analyses based on the Fokker-Planck equation, while our nonharmonic cases feature a double-well potential and an isotropic quartic potential.  In particular, we report two simple methods that can help understand gyrating patterns.  For harmonic potentials, we use the Fokker-Planck equation to survey the NESS dynamical characteristics, i.e., the NESS currents gyrate along the equiprobability contours and the stationary point of flow coincides with the potential minimum.  As a contrast, the NESS results in our nonharmonic potentials show that these properties are largely absent, as the gyrating patterns are much distinct from those of corresponding probability distributions.  Furthermore, we observe a critical case of the double-well potential, where the harmonic contribution to the gyrating pattern becomes absent, and the NESS currents do not circulate about the equiprobability contours nearby the potential minima even at low temperatures.
\end{abstract}

%\pacs{}
\maketitle

%%%%%%%%%%%%%%%%%%%%%%%%%%%%%%%%%%%%%%%%%%%%%%%%%%%%%
\section{Introduction}
\label{sec_intro}
The recent discussions over autonomous Brownian gyrators\cite{Filliger_PRL2007, Imparato_gyrator, quantum_gyrator, harmonic_analytic2013, harmonic_analytic2018, RC_gyrator_2017, exp_gyrator2017} have shed light on the development of autonomous Brownian engines\cite{Buttiker, Landauer, Reimann02, Park16, Eshuis10, Hartmann15, Serra-Garcia16, Gonzalez19, Stolovitzky98, Ryabov16, Holubec17, Kalinay18, He20}.  In the simplest description of such systems, a Brownian particle performs two-dimensional random walks under the influence of some conservative potential, while thermal fluctuations of unequal strength are supplied along the two cartesian coordinates, respectively.  The resulting dynamics is signatured by an average nonzero, circulating flow field.
The famous Feynman's ratchet problem \cite{Feynman63} can be considered as a particular example of the autonomous Brownian gyrators\cite{Stolovitzky98, Ryabov16, Holubec17, Kalinay18, He20}.  In the ratchet system, one of the coordinates is periodic, while the random walker is subject to a nonlinear potential, and the average dynamics exhibits unidirectional movement along the periodic coordinate.
The autonomous gyrating property makes it a sought-after candidate for the realization of microscopic heat engines.

Experimentally, realization of a Brownian gyrator was demonstrated through a colloidal system using optical tweezers along with artificial noises\cite{exp_gyrator2017}.  Meanwhile, it has been noticed that a coupled RC circuit system with thermal noises can serve as a complete analog to a Brownian gyrator under a harmonic potential\cite{RC_gyrator_2017, Ciliberto13}.  Similar mechanical and electrical realizations of these autonomous engines with the use of artificial noises were reported and often featured with less trivial interactions \cite{Eshuis10,Hartmann15,Serra-Garcia16,Gonzalez19, He20}.

%The original proposal of autonomous Brownian gyrators\cite{Filliger_PRL2007} and subsequent demonstrations\cite{} were mostly studied utilizing the template of harmonic potentials.  Thanks to the harmonic structure, their nonequilibrium steady-state (NESS) dynamics have been analyzed thoroughly in previous studies \cite{Ao_PNAS2005, harmonic_analytic2013, harmonic_analytic2018, RC_gyrator_2017}.
Theory-wise, the nonequilibrium steady-state (NESS) dynamics for harmonic potentials have been studied quite thoroughly\cite{Ao_PNAS2005, Imparato_gyrator, harmonic_analytic2013, harmonic_analytic2018, Park16, RC_gyrator_2017}.  In particular, these studies brought about the important features that the NESS currents gyrate about the equiprobability contours, and the stationary points of flow coincide with the potential minima.  These successes in harmonic systems also lead to a keen desire in the investigation over a broader range of potentials.
On the one hand, one should ask whether these traits from harmonic results still persist regarding more general potential cases.  On the other hand, the nonlinearity in potentials often brings novelty to the NESS dynamics.
%{\color{red} For example, Feynman's ratchet problem can be treated as a two-dimensional gyrating system subject to a nonlinear potential\cite{Stolovitzky98, Ryabov16, Holubec17, Kalinay18}. }
For example, it has been mentioned in Ref.~\cite{Ao_PNAS2005} that the anharmonicity in potentials can act as ``external'' currents that shifts the stationary point of flow for the correspondingly harmonic problems.  And therefore, unlike the results of harmonic potentials, the overall currents do not necessarily gyrate about the potential contours.  Furthermore, new gyrating patterns may emerge due to the existence of anharmonicity\cite{Kwon_PRE2011}.

While there have been numerous theoretical efforts towards the general treatment in NESS dynamics\cite{Risken, Derrida2001, Sasa2006, Ao2007, Seifert2012, Noh_2014}, we choose to focus our current work on a two-dimensional Brownian system under some conservative potential, due to the desire to understand the gyrator characteristics.
%The goal of our current work is to study the gyrating characteristics and its relation to the potential and corresponding probability distribution.
The paper is organized as follows.  First we introduce in Sec.~\ref{sec_modelmethod} our system of interest, and its corresponding Fokker-Planck equation.  In Sec.~\ref{sec_circulation} we introduce two simple criteria that can examine the gyrating directions given a general potential.  %In particular, one of these methods relies on the use of the second law of thermodynamics over an infinitesimal loop.
To study the gyrating characteristics, we first re-derive in Sec.~\ref{sec_harmonic} the NESS dynamics for the harmonic potential, starting from its major characteristic that the NESS currents gyrate about the equiprobability contours.
As for nonharmonic potentials, we consider two specific examples: a double-well potential is studied in Sec.~\ref{sec_double_well}, and an isotropic quartic potential is studied in Sec.~\ref{sec_iso_quartic}.  In Sec.~\ref{sec_dissipation} we briefly discuss about the entropy change along the autonomous trajectories.
%As to the study of Brownian gyrators, it is important to ask if there exists an efficient method that can predict the gyrating patterns considering anharmonic or even nonharmonic potentials.
%For example, can the harmonic treatment always be valid nearby the bottom of a potential well?
%In thermal equilibrium, considering the behavior near the bottom of a general potential, the harmonic approximation can often lead to good results both qualitatively and quantatively.  On the other hand, the use of an analogous methodology is yet to be clarified regarding the study of nonequilibrium steady states (NESS).
%Kwon, Ao, and Thouless\cite{PNAS_Ao} has demonstrated a recipe that helps analyze the structure and dynamics of a random walker under the influence of linear forces.  In their pivotal work, they also discussed about the possible existence of other NESS solutions that may occur due to the existence of nonlinear forces.

\section{System}
\label{sec_modelmethod}
We consider a Brownian particle in two dimensions. The corresponding Langevin equation is
\begin{equation}
  \gamma \dot{\vec{r}} = - \nabla U + {\vec{\xi}} \, ,
  \label{eqn_Langevin}
\end{equation}
where $\vec{r} = \left(\begin{array}{c} x_1 \\ x_2 \end{array} \right)$ is the position of the particle, $U$ is the potential energy, and $\vec{\xi}$ is the random force.  The components of the random force are Gaussian white and uncorrelated, namely, $\langle \xi_i(t)\xi_j(t')\rangle=2\gamma k_B T_i \delta_{ij}\delta(t-t')$, where $k_B$ is the Boltzmann constant.  Throughout this report we only consider the cases where $T_1 > T_2$.  Therefore the system does not achieve thermal equilibrium, and our focus is on its NESS characteristics.
	
For simplicity, we adopt the dimensionless convention that $k_B =1$ and $\gamma=1$ in our analytical work.  The latter can be achieved through a rescale of time by the factor $1/\gamma$.  The probability distribution function of this system $P$ is described by the Fokker-Planck equation:
\begin{equation}
  \frac{\partial P}{\partial t}  = \nabla \cdot \hat{\bf D} \nabla P + \nabla \cdot (P \nabla U)\, ,
  \label{eqn_FP}
\end{equation}
where $\displaystyle \hat{\bf D} = \left(\begin{array}{cc} T_1 & 0 \\ 0 & T_2 \end{array} \right)$ under our dimensionless description.  Since we are interested in the NESS only, the left-hand side of Eq.~\ref{eqn_FP} drops to zero.  The flux density $\vec{J}$ is defined as
\begin{equation}
  \vec{J}  \equiv - \hat{\bf D} \nabla P - P  \nabla U\, .
  \label{eqn_J1}
\end{equation}
For convenience let us denote $\phi \equiv -\log P$.  Then one can rewrite the flux density as
\begin{equation}
  \vec{J}  = P (\hat{\bf D} \nabla \phi -  \nabla U ) \equiv P \vec{v}_{\rm av}\, ,
  \label{eqn_J2}
\end{equation}
where $\vec{v}_{\rm av}$ represents the average velocity of the probability flux.  The equation of continuity at NESS then gives
\begin{equation}
  \nabla \cdot \vec{J} = P \nabla \cdot \vec{v}_{\rm av} + \vec{v}_{\rm av} \cdot \nabla P = 0\, ,
  \label{eqn_cont1}	
\end{equation}
or equivalently,
\begin{equation}
  \nabla \cdot \vec{v}_{\rm av} - \vec{v}_{\rm av} \cdot \nabla \phi = 0\, .
  \label{eqn_cont2}
\end{equation}

%Our previous study of an electrical autonomous Brownian gyrator...

In this work we define a NESS flow path to be that following the movement of $\vec{v}_{\rm av}$.
Let us assume that the NESS currents gyrate about the equiprobability contours, i.e., $\vec{v}_{\rm av} \cdot \nabla \phi =0$.  Then Eq.~\ref{eqn_cont2} shows that this is equivalent to the statement $\nabla \cdot \vec{v}_{\rm av} =0$.  Therefore, the NESS flow field can be compared to that of an ideal fluid, and $\displaystyle\frac{dP}{dt}= 0$ along any NESS flow path (it can also be understood knowing that the NESS flow path coincides with some equiprobability contour).  Now we consider a stationary point of $U$, i.e., $\nabla U =\vec{0}$\cite{footnote}.  Then Eq.~\ref{eqn_J2} implies that $\vec{v}_{\rm av} = \hat{\bf D} \nabla \phi$.  Since $\hat{\bf D}$ is positive, the assumption that $\vec{v}_{\rm av} \cdot \nabla \phi =0$ requires that this point is also a stationary point of $\phi$, and therefore $\vec{v}_{\rm av}$ must vanish at this position.  The inverse statement is also true: nearby a stationary point of $\phi$, since the NESS currents gyrate about the equiprobability contours, there always exist currents of different directions in a neighborhood about the stationary point of $\phi$.  As the neighborhood of consideration approaches infinitesimal, this property can only be satisfied by the requirement that $\vec{v}_{\rm av}=\vec{0}$, i.e., the stationary point of $\phi$ must be a stationary point of flow, and from Eq.~\ref{eqn_J2} one deduces that $U$ is also a stationary point.  From these discussions, one can conclude that if the property $\vec{v}_{\rm av}\cdot \nabla \phi = 0$ holds, then the stationary points of $U$ and $\phi$ and the stationary point of flow must coincide with each other.

%In this current study, we demonstrate through analytical studies and simulations that while the aforementioned properties are possessed in systems of harmonic potentials, they are largely absent when the potential is nonharmonic.
While the aforementioned properties are valid in systems of harmonic potentials, they are largely absent when the potential becomes nonharmonic.  In this study we investigate their gyrating patterns through simulations and analytical arguments.  We perform our simulations via integrations over Eq.~\ref{eqn_Langevin} through discretization following It\^o's scheme (using the Euler method).  In our simulations, we set $\gamma=9 \times 10^{-4}$ and choose the discretized time interval $\Delta t = 10^{-5}$.  This is equivalent to our setting of analytical work $\gamma =1$ with the effective discrete time interval $\Delta t =0.1/9$.  For each model the simulation is performed over $10^7$ iterations.  The recorded positions in $x_1$ and $x_2$ are sorted into bins of width $x_{\rm bin}$, so that statistical distributions can be obtained. The flux density $\vec{J}$ is derived via the formula $\vec{J} = P\vec{v}_{\rm av}$, while $\vec{v}_{\rm av}$ is obtained by
\begin{eqnarray}
 \vec{v}_{\rm av}(\vec{r}) &\approx& \frac{1}{2\Delta t} \{ \langle [\vec{r}(t+\Delta t)-\vec{r}(t)] |\vec{r}(t)=\vec{r} \rangle \nonumber \\ && \,\, + \langle [\vec{r}(t)-\vec{r}(t-\Delta t)] |\vec{r}(t)=\vec{r}\rangle \} \, ,
 \label{eqn_v_ave}
\end{eqnarray}
i.e., the NESS velocity at the grid $\vec{r}$ is computed by the average of all the discrete transitions that either start or end at $\vec{r}$.  In practice Eq.~\ref{eqn_v_ave} is computed using the coarse-grained spatial coordinates.

%(It can be shown that if the potential is harmonic, the NESS flows gyrate about the equiprobability contours, and therefore $\nabla \cdot \vec{v}_{rm av} =0$ everywhere.)
%%%%%%%%%%%%%%
\section{Direction of circulation}
\label{sec_circulation}
In this work we employ two methods to study the direction of circulation in the NESS flow field.  First, we consider the curl of the average velocity, as from Eq.~\ref{eqn_J2} one can derive
\begin{eqnarray}
 \nabla \times \vec{v}_{\rm av} &=& \nabla \times \hat{\bf D} \nabla \phi -\nabla \times \nabla U \nonumber \\ &=& - (T_1-T_2)\partial_1 \partial_2 \phi \, .
 \label{eqn_curl_v1}
\end{eqnarray}
Let us define the tilted axes
\begin{eqnarray}
 x'_1 &\equiv & \frac{x_1+x_2}{\sqrt{2}}\, ,
 \label{eqn_prime1}
\\
 x'_2 &\equiv & \frac{x_1-x_2}{\sqrt{2}}\, ,
 \label{eqn_prime2}
\end{eqnarray}
while
\begin{eqnarray}
 \partial'_1 &\equiv & \frac{\partial}{\partial x'_1} = \frac{\partial_1 + \partial_2}{\sqrt{2}} \, ,
 \label{eqn_prime3}
\\
 \partial'_2 &\equiv & \frac{\partial}{\partial x'_2} = \frac{\partial_1 - \partial_2}{\sqrt{2}} \, .
 \label{eqn_prime4}
\end{eqnarray}
Then Eq.~\ref{eqn_curl_v1} becomes
\begin{equation}
 \nabla \times \vec{v}_{\rm av} = - \frac12 (T_1-T_2)(\partial'^2_1 - \partial'^2_2) \phi \, .
 \label{eqn_curl_v2}
\end{equation}
Equation~\ref{eqn_curl_v2} tells us that the circulation of the NESS currents is correlated with the difference in curvature of $\phi$ along the directions $x'_1$ and $x'_2$.  Since $T_1>T_2$ in all our studied cases, one can observe a clockwise gyration ($\nabla \times \vec{v}_{\rm av}$ being negative) if $\partial_1 \partial_2 \phi > 0$ (i.e., $\partial'^2_1 \phi > \partial'^2_2 \phi$) and vice versa.

Alternatively, the gyrating direction of the NESS currents can be studied utilizing the second law of thermodynamics.  Let us consider a NESS flow cycle on the $x_1-x_2$ plane.  In our overdamped system as described by Eq.~\ref{eqn_Langevin}, the amount of heat dissipating into the reservoir $T_1$ during the cycle is $Q_{1,{\rm ssf}} = -  \oint_{\rm ssf} \partial_1 U dx_1  $, and the corresponding heat dissipating into reservoir $T_2$ is $Q_{2,{\rm ssf}} = - \oint_{\rm ssf} \partial_2 U dx_2$.  Note that we use the subscript ``ssf'' to designate that the physical quantity is evaluated along a NESS flow path, where an infinitesimal displacement can be understood by $d\vec{r}= (dx_1, dx_2) = \vec{v}_{\rm av} dt$\cite{Sekimoto_book}.  The fact that $\oint_{\rm ssf} \nabla U \cdot d\vec{r} = 0$ reminds us that $ Q_{1,{\rm ssf}} =- Q_{2,{\rm ssf}} $.  The total entropy change of an infinitesimal path is $dS_{\rm tot}  = dS_{\rm sys} + dS_{Q} $, where $S_{\rm sys} \equiv - \log P = \phi$ is the system entropy and $dS_{Q}= dQ_1/T_1 +dQ_2/T_2$ is the net entropy change of the heat reservoirs.  Since the system entropy is a state function, its change over any closed path is equal to zero.
%Consider the probability flux distribution on the $x_1-x_2$ plane.  The paths following the flowing directions represent NESS trajectories, and
Therefore the total entropy change over a closed path is

\begin{eqnarray}
 \oint_{\rm ssf} dS_{\rm tot} &=& \oint_{\rm ssf}  dS_{Q} \nonumber \\&=& \left(\frac{1}{T_2} - \frac{1}{T_1}\right) \oint_{\rm ssf} \partial_1 U dx_1  >0 \, ,
 \label{eqn_entropy_cycle}
\end{eqnarray}
as is required by the second law of thermodynamics (in Sec.~\ref{sec_dissipation} we show that the change in total entropy is positive along any NESS flow path, which stands as a version of the second law).  The sign in Eq.~\ref{eqn_entropy_cycle} indicates that regarding the NESS probability flux distribution, only the cycles with $\oint_{\rm ssf} \partial_1 U dx_1 >0 $ are legitimate.

%Now one considers an infinitesimal rectangular area of width $\delta x$ and height $\delta y$.  By integrating over the counterclockwise loop, one can derive
%\begin{eqnarray}
% - \oint_{\rm CCW} \partial_1 U dx_1 & \approx & \delta x \delta y \cdot \partial_1\partial_2 U(x,y) \nonumber \\
% &=& \frac{\delta x \delta y}{2} (\partial'^2_1 - \partial'^2_2)U \, .
% \label{eqn_gyration}
%\end{eqnarray}
%From equations \ref{eqn_entropy_cycle} and \ref{eqn_gyration} one can conclude that the autonomous gyration is clockwise if $\partial_1\partial_2 U >0$ (or equivalently, $\partial'^2_1 U > \partial'^2_2 U $ ) and vice versa.

Using the Stokes' theorem, one can rewrite Eq.~\ref{eqn_entropy_cycle} as
\begin{eqnarray}
\oint_{\rm ssf} dS_{\rm tot}  &=& - \oint_{\rm ssf} (\hat{\bf D}^{-1} \nabla U) \cdot d\vec{r} \nonumber \\ &=& - \int_{\rm ssf} \nabla \times (\hat{\bf D}^{-1} \nabla U)\cdot d\vec{A} \, \, .
 \label{eqn_entropy_cycle2}
\end{eqnarray}
%where the directions of the length and area differential elements $d\vec{l}$ and $d\vec{A}$ are chosen such that $\oint_{\rm ssf} dS_{\rm tot}$ is positive.
%where the direction of the loop integral is chosen such that the integral is positive.
%From Eq.~\ref{eqn_entropy_cycle2} one finds that $- \left< \oint (\hat{\bf D}^{-1} \nabla U) \cdot d\vec{l} \right> >0$ along an average gyrating trajectory.
The expressions in Eq.~\ref{eqn_entropy_cycle2} lead us to speculate that one can utilize the second law on infinitesimal cycles as well, i.e., the expression $- \nabla \times (\hat{\bf D}^{-1} \nabla U)$ can serve as an indicator of the local gyrating trend.  In the next paragraph, we shall start from a different approach and show that this differential expression gives information about the local gyration.  Since $\nabla \times (\hat{\bf D}^{-1} \nabla U) = (T_1-T_2)\partial_1 \partial_2 U/(T_1 T_2)$, one can therefore deduce that the local gyrating direction is clockwise if $\partial_1\partial_2 U >0$ (or equivalently, $\partial'^2_1 U > \partial'^2_2 U $ ) and vice versa.

It is worth noting that one can rescale the variable $x_1$ such that the system behaved as the one with identical temperatures on both dimensions: let $y_1 \equiv x_1/c$ and $y_2 \equiv x_2$.  Then one can rewrite Eq.~\ref{eqn_Langevin} as
%\begin{eqnarray}
% \dot{y_1} &=& g_{1} + \eta_1 \nonumber \\
% \dot{y_2} &=& g_{2} +\eta_2 \, ,
% \label{eqn_Langevin2}
%\end{eqnarray}
\begin{equation}
 \dot{\vec{y}}= \vec{f}_y +\vec{\xi}_y \, ,	
 \label{eqn_Langevin2}
\end{equation}	
where $\xi_{y,1} \equiv \xi_1/c$ and $\xi_{y,2} = \xi_2$, while $f_{y,1}=-\partial_1 U/c$ and $f_{y,2}=-\partial_2 U$.  In the following we use the subscript $y$ to denote physical quantities in the ($y_1, y_2$) coordinates.  Through the choice $c= \sqrt{T_1/T_2}$, the random variables $\xi_{y,1}$ and $\xi_{y,2}$ can have identical distributions.  Note that in this coordinate system the force $\vec{f}_y$ in Eq.~\ref{eqn_Langevin2} is not conservative.  The NESS velocity under this rescaled coordinates is
\begin{equation}
 \vec{v}_{y, \rm av} = \hat{\bf D}_y \nabla_y \phi_y + \vec{f}_y  \, .
\end{equation}
Since the heat baths along the two dimensions are now identical, $\hat{\bf D}_y$ is proportional to the identity matrix, and therefore
\begin{eqnarray}
 \nabla_y \times \vec{v}_{y, \rm av} &=& \nabla_y \times \vec{f}_y \nonumber \\
 &=& \partial_{y,1} f_{y,2}-\partial_{y,2} f_{y,1} \nonumber \\
 &=& c\,\partial_1 f_2 - \frac{1}{c}\,\partial_2 f_1 \nonumber \\
 &=& -\frac{1}{c} (c^2-1) \partial_1\partial_2 U \nonumber \\
 &=& -\sqrt{T_1T_2} (\nabla \times \hat{\bf D}^{-1}\nabla U) \, .
 \label{eqn_curl_vy}
\end{eqnarray}
The last line in Eq.~\ref{eqn_curl_vy} shows that our speculation of local gyrating direction from the total entropy argument is evidenced through the discussion over the curl of the average velocity in the rescaled coordinates.

The difference between these two methods concerning the study of gyrating directions is mainly attributed to the uses of $\phi$ and $U$, respectively.  Despite this difference, their qualitative results are mostly similar due to the geometric resemblance between $\phi$ and $U$.

\section{Harmonic potential}
\label{sec_harmonic}
We first revisit the harmonic potential cases through the analysis based on its Fokker-Planck equation.
Note that the symmetry of the harmonic potential, namely $U(-\vec{r}) = U(\vec{r})$, leads to $\phi(-\vec{r})=\phi(\vec{r})$, since the random walker cannot sense any physical difference between any opposite pair of points $\vec{r}$ and $-\vec{r}$.  In this case, Eqs.~\ref{eqn_J1} and \ref{eqn_J2} indicates that $\vec{J}(-\vec{r}) = -\vec{J}(\vec{r})$ and $\vec{v}_{\rm av}(-\vec{r}) = -\vec{v}_{\rm av}(\vec{r})$.  Therefore, $\vec{v}_{\rm av}=\vec{0}$ at the origin, and from Eq.~\ref{eqn_cont2} one also finds that $\nabla \cdot \vec{v}_{\rm av} =0$ at the origin.

In our previous study\cite{RC_gyrator_2017} of an autonomous Brownian gyrator demonstrated by a coupled RC system, the potential profile is quadratic, and the resulting $\phi$ also possesses a quadratic form.  For general random walks over two dimensions under harmonic potentials, one can also use the quadratic ansatz in $\phi$.  Following Eq.~\ref{eqn_J2}, one finds that $\vec{v}_{\rm av}$ is linear in $x_1$ and $x_2$, and therefore $\nabla \cdot \vec{v}_{\rm av}$ is a constant.
Our discussion in the last paragraph concludes that this constant is zero, and the NESS currents gyrate around the equiprobability contour lines.

Since $\vec{v}_{\rm av}$ is perpendicular to $\nabla P$ and therefore $\nabla \phi$, one can rewrite $\vec{v}_{\rm av} =  - \epsilon \hat{\bf Y} \nabla \phi$, where
\begin{equation}
\hat{\bf Y} \equiv \left(\begin{array}{cc} 0 & 1 \\ -1 & 0 \end{array} \right)
\end{equation}
is a $2\times2$ matrix that rotates $\nabla \phi$ clockwise by the angle $\pi/2$, and $\epsilon$ is a scalar function to be determined.
%(for a harmonic potential one can proceed to show that $\epsilon$ is a constant).
Then Eq.~\ref{eqn_J2} becomes
\begin{equation}
  \vec{v}_{\rm av} = \hat{\bf D} \nabla \phi -  \nabla U = - \epsilon \hat{\bf Y} \nabla \phi
  \label{eqn_v}
\end{equation}
or
\begin{equation}
  (\hat{\bf D} + \epsilon \hat{\bf Y}) \nabla \phi = \nabla U \, .
  \label{eqn_AoPing}
\end{equation}
Note that $\hat{\bf D} + \epsilon \hat{\bf Y}$ is equivalent to the matrix $\hat{\bf G}$ as described in Refs.\cite{Ao_PNAS2005}.  %Meanwhile, Eq.~\ref{eqn_AoPing} also implies that the extrema of $U$ and $\phi$ coincide.

Substituting Eq.~\ref{eqn_v} into the relation $\nabla \cdot \vec{v}_{\rm av} = 0$, one gets (using the fact that $\vec{a} \cdot \hat{\bf Y} \vec{b} = \vec{a} \times \vec{b} \equiv a_1 b_2-a_2 b_1$ for any $\vec{a}$ and $\vec{b}$)
\begin{equation}
  \nabla \epsilon \times \nabla \phi +\epsilon \nabla \times \nabla \phi = \nabla \epsilon \times \nabla \phi = 0 \, .
\end{equation}
Therefore either $\nabla \epsilon$ is parallel to $\nabla \phi$, or $\epsilon$ is simply a constant.

From Eq.~\ref{eqn_AoPing} one has
\begin{eqnarray}
  0 &=& \nabla \times \nabla \phi = \nabla \cdot \hat{\bf Y} \hat{\bf G}^{-1} \nabla U \nonumber \\ &=& \nabla \cdot \frac{\hat{\bf Y} ( \hat{\bf D}_{\rm a}- \epsilon \hat{\bf Y}) \nabla U}{\det \hat{\bf G}} \nonumber \\ &=& -\frac{2\epsilon \nabla \epsilon}{\det ^2 \hat{\bf G}} \times (\hat{\bf D}_{\rm a} - \epsilon \hat{\bf Y}) \nabla U \nonumber \\ &&\ \ \ \ + \frac{\nabla \times (\hat{\bf D}_{\rm a}\nabla U) + \nabla \cdot (\epsilon \nabla U)}{\det \hat{\bf G}} \, ,
  \label{eqn_curl_phi1}
\end{eqnarray}	
where $\hat{\bf D}_{\rm a}$ is the adjugate matrix of $\hat{\bf D}$.  The first term of Eq.~\ref{eqn_curl_phi1} is 0, and therefore one has
\begin{equation}
  - \nabla \times (\hat{\bf D}_{\rm a}\nabla U) = \nabla \epsilon \cdot \nabla U  + \epsilon \nabla \cdot \nabla U \, .
  \label{eqn_curl_phi2}
\end{equation}
Since the potential is quadratic, the solution that $\epsilon$ is a constant serves as a legitimate answer for the above equation, as
\begin{equation}
  \epsilon = -\frac{\nabla \times (\hat{\bf D}_{\rm a} \nabla U)}{\nabla^2 U} = -(T_1-T_2)\frac{\partial_1 \partial_2 U}{\nabla^2 U}\, .
  \label{eqn_epsilon}
\end{equation}
From Eq.~\ref{eqn_AoPing}, one sees that the result that $\epsilon$ is a constant leads to a quadratic $\phi$.  This is consistent with our initial conjecture.  Using the expressions $\phi \equiv \vec{r} \cdot \hat{\bf O} \cdot \vec{r}$ and $U \equiv \vec{r} \cdot \hat{\bf U} \cdot \vec{r}$, one can derive from Eq.~\ref{eqn_AoPing} that $(\hat{\bf D} + \epsilon \hat{\bf Y}) \hat{\bf O} = \hat{\bf U}$, and therefore
\begin{equation}
  \hat{\bf O} = (\hat{\bf D} + \epsilon \hat{\bf Y})^{-1} \hat{\bf U} \, .
  \label{eqn_O}
\end{equation}	
If one substitutes the result of $\phi$ into Eq.~\ref{eqn_curl_v1}, it is straightforward to show that
\begin{equation}
  \nabla \times \vec{v}_{\rm av} = -\frac{2({T_1}^2-{T_2}^2)}{T_1 T_2+\epsilon^2}U_{12} \, ,
  \label{eqn_curl_v3}
\end{equation}
where $U_{12}$ is the off-diagonal element of the matrix $\hat{\bf U}$.  Because $T_1>T_2$, the average autonomous gyration is counterclockwise if $U_{12}<0$ and vice versa.  Therefore, for the case of harmonic potential, both of our methods regarding the circulating direction are in full agreement.

\begin{figure}[h]
 \centering
 \includegraphics[width=0.5\textwidth]{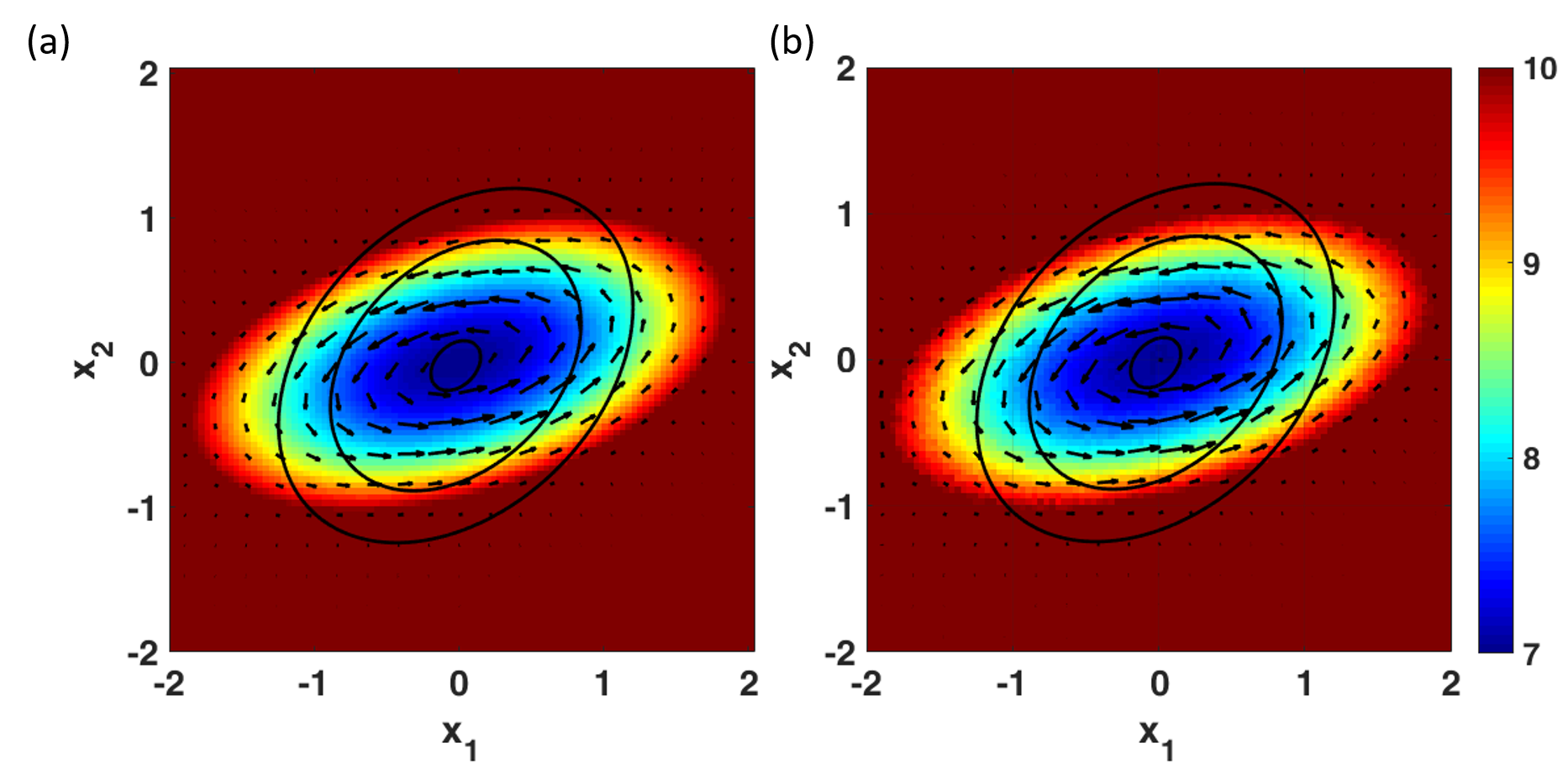}
 \caption{(a) Theoretical prediction and (b) numerical simulation result of a Brownian gyrator under the harmonic potential $\displaystyle U = \frac12 k ({x_1}^2 + {x_2}^2) + k' x_1 x_2$.  We use the setting $k=1.5$ and $k'=-0.5$, while $T_1 = 0.8$ and $T_2 = 0.2$.  The closed loops represent equipotential contours, while the color maps and vectors designate the NESS distribution of $\phi = -\log P$ and flux densities ${\vec{J}}$, respectively.  The distribution of $\phi$ is derived using the bin size $x_{\rm bin}=4/99$ (denominator representing number of grids considered in each dimension), while $x_{\rm bin}=4/19$ is adopted in the calculation of $\vec{J}$.  }
 \label{fig_harmonic}	
\end{figure}

We use the harmonic potential $\displaystyle U(\vec{r}) = \frac{k}{2} ({x_1}^2 +{x_2}^2) + k' x_1 x_2$ as a specific example, and we apply the parameter settings $k=1.5$ and $k'=-0.5$, while $T_1=0.8$ and $T_2=0.2$.  Our analytical and numerical results are presented in Fig.~\ref{fig_harmonic}(a) and (b), respectively.  Our simulation result again exhibits the feature that the NESS currents gyrate along the equiprobability contours.  Moreover, the counterclockwise gyrating direction echoed the prediction from Eq.~\ref{eqn_curl_v3}.  Alternatively, the circulating direction can be hinted through the comparison of the curvatures of $U$ along the $x'_1$ and $x'_2$ axes.  One can observe in Fig.~\ref{fig_harmonic} that the curvature along the $x'_1$ direction is milder than that along the $x'_2$ direction, i.e., $\partial'^2_1 U < \partial'^2_2 U$.  Follow the argument in Eqs.~\ref{eqn_entropy_cycle} and \ref{eqn_entropy_cycle2}, one also reaches the conclusion of a counterclockwise gyration.

\section{Double-well potential}
\label{sec_double_well}
\begin{figure*}[t]
 \centering
 \includegraphics[width=\textwidth]{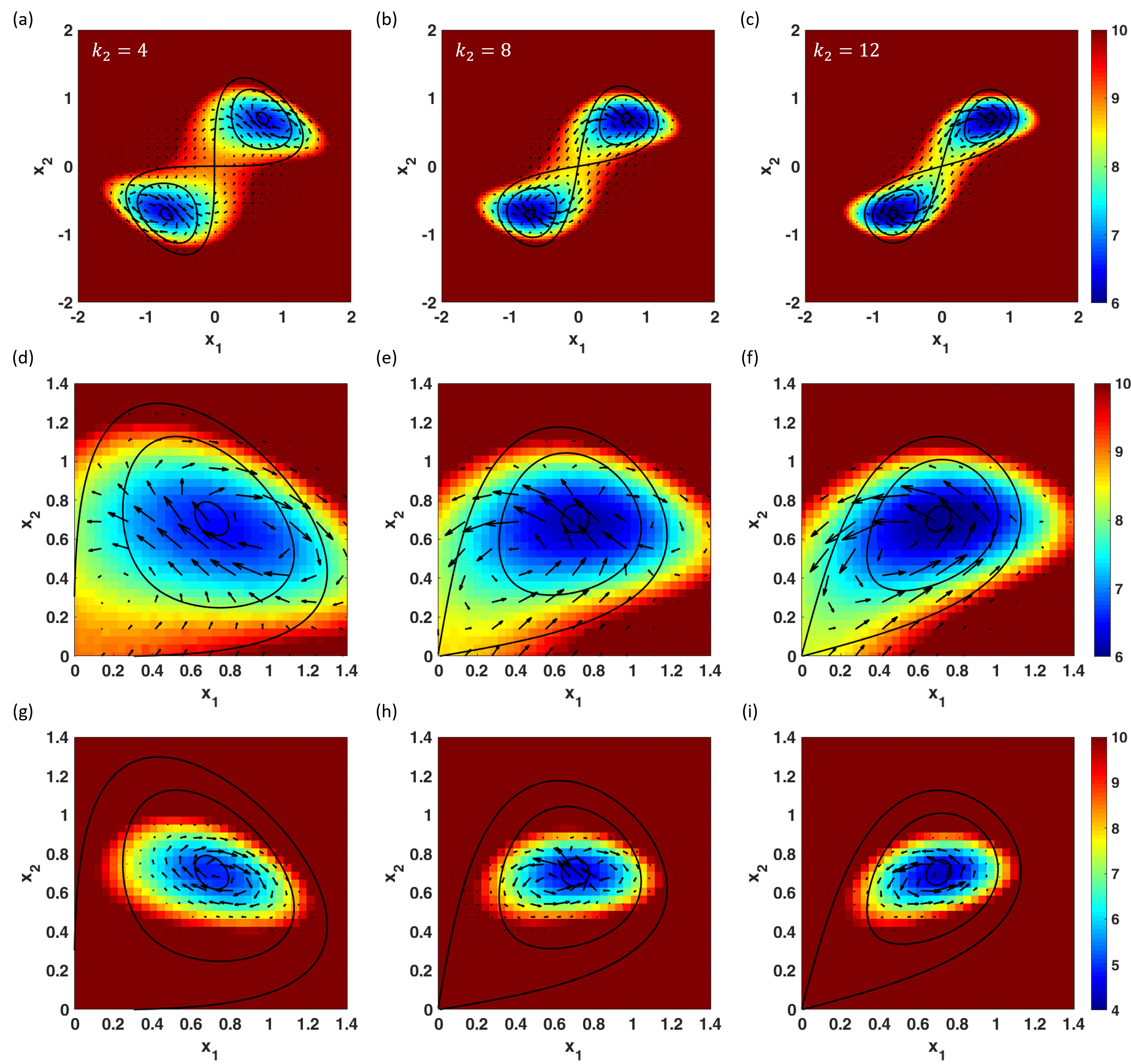}
 \caption{NESS simulation results of the double-well potential $U = {x'_1}^4-2 {x'_1}^2 +\frac12 k_2 {x'_2}^2$.  The values $k_2=4, 8, 12$ are applied in the figures of the first, second, and third columns, respectively.  The temperatures $T_1 = 0.8$ and $T_2=0.2$ are applied in the first and second rows, while the third row provides the result of our low-temperature setup ($T_1=0.16$ and $T_2=0.04$).  The notational designation follows that of Fig.~\ref{fig_harmonic}.  The distribution of $\phi$ is derived using the bin size $x_{\rm bin}=4/99$.  As to the calculation of $\vec{J}$, we adopt the bin size $x_{\rm bin}=4/29$ in the high-temperature setup (first and second rows) and $x_{\rm bin}=4/58$ in the low-temperature setup (third row). }
 \label{fig_double_well}
\end{figure*}

Next we study the autonomous Brownian gyrators under nonharmonic potentials.  We first consider a double-well potential  (using the primed coordinates as defined in Eqs.~\ref{eqn_prime1} and \ref{eqn_prime2})
\begin{equation}
  U(\vec{r}) = {x'_1}^4-2 {x'_1}^2 +\frac12 k_2 {x'_2}^2 \, .
  \label{eqn_double_well}
\end{equation}
Note that the two wells are lined up in the $x'_1$ axis.  The potential is harmonic along the $x'_2$ axis and asymptotically quartic along the $x'_1$ axis.  Our simulation results of the NESS probability and flux distributions are presented in Fig.~\ref{fig_double_well} using various values of $k_2$.  In the first and second rows of Fig.~\ref{fig_double_well} we consider the temperature setting $T_1=0.8$ and $T_2=0.2$, while in the third row we use $T_1=0.16$ and $T_2=0.04$.  First, one can find significant areas in the first and second rows of Fig.~\ref{fig_double_well} that the NESS currents do not circulate about the equiprobability contours, i.e., $\vec{v}_{\rm av} \cdot \nabla P \neq 0$.  Therefore, the system entropy is not a constant of motion along the NESS flow paths.  Instead, it exhibits some sort of oscillatory pattern during a cycle.  Moreover, one can observe in the first second rows of Fig.~\ref{fig_double_well} nonvanishing NESS currents at the probability maxima, and the probability and potential extrema do not coincide (the second row of Fig.~\ref{fig_double_well} provides a closer look).

%%%While the double-well potential is approximately harmonic if one stays close to a local minimum, it is natural to wonder that the NESS result will approach the harmonic behavior at low temperature, where the random walker cannot easily access the nonharmonic region.  However, as we lower the temperatures along $x_1$ and $x_2$ to $T_1=0.16$ and $T_2=0.04$, respectively, the simulation result nearby a potential minimum still does not exhibit the harmonic features, and one can observe a significant NESS flow at the probability maxima (see Fig.~\ref{fig_double_well}(d)).  Note that at such a low-temperature regime, the random walker can hardly jump across the potential barrier.

%For our observed cases, the potential and probability extrema are very close.  Yet the resulting NESS current at those postitions are remarkably significant, which is seemingly counterintuitive.  Considering Eq.~\ref{eqn_J1}, one may be tempted to conclude that at one probability maximum (or similarly potential minimum), one gradient term disappears while the magnitude of the other term has to be mild, and the resulting current must be less significant.  On the other hand, a more careful examination on Eq.~\ref{eqn_J2} reveals that the fast exponential decay in $P$ outweighs the growth of $\vec{v}_{rm av}$ as one leaves a probability maximum.
In addition to the fact that the potential energy minima are not stationary points of flow, Fig.~\ref{fig_double_well} also reveals that there exist two stationary points of flow near each potential minimum.  Furthermore, the gyrating directions around the two stationary points of flow are opposite to each other.  While this result is due to nonharmonic effects, it can be understood using Eq.~\ref{eqn_curl_v2}.  First, the probability distribution shows that $\phi$ exhibits a similar double-well shape compared with the potential profile.  If $\phi$ has a geometric structure that is similar to the potential, then along the $x'_1$ axis, the curl of $\vec{v}_{\rm av}$ must be negative at large $x'_1$.  This is because the quartic-like behavior along the $x'_1$ axis results in a larger curvature that outweighs the rather harmonic behavior along the $x'_2$ axis.  Thus this argument leads to a clockwise gyrating behavior at large $x'_1$.  Moreover, for this double potential one has $\partial'^2_1 U - \partial'^2_2 U = 12 {x'_1}^2 - 4 - k_2$.  If we follow the argument from the total entropy production following Eqs.~\ref{eqn_entropy_cycle} and \ref{eqn_entropy_cycle2}, we find the argument predicts a clockwise gyration for $|x'_1| > \sqrt{(k_2+4)/12}$ and a counterclockwise one for $|x'_1|<\sqrt{(k_2+4)/12}$.  In particular, the potential energy minimum is located at the clockwise gyrating region for $k_2=4$ and counterclockwise gyrating region for $k_2$=12, which are evidenced by our simulation result in Fig.~\ref{fig_double_well}.

Note that the setting $k_2=8$ stands as a special case in our discussion.  First, the above analysis on the gyrating behavior holds for all temperatures that meet the criterion $T_1>T_2$.  And since for $k_2=8$ the potential minima occur at the positions $(x'_1=\pm 1, x'_2=0)$, one can always anticipate two oppositely gyrating regions neighboring a potential minimum.  Our simulation result also confirms this gyrating signature, while this behavior can still be observed even with our low-temperature setup ($T_1=0.16$ and $T_2=0.04$; please refer to the third row of Fig.~\ref{fig_double_well}).  At a first look, this result appears perplexing, since unlike the results for other values of $k_2$, it does not approach the harmonic behavior at low temperatures.  This can be understood knowing that for the case $k_2=8$, the shape of the potential nearby each minimum is harmonic but circular (the latter fact can be shown by $\partial'^2_1 U - \partial'^2_2 U =0$ at the potential minimum).  As a result, the leading harmonic approximation gives no NESS currents\cite{Filliger_PRL2007}, while the nonharmonic part of the potential makes the dominant contribution.

There is a similar puzzle regarding the NESS probability distribution that needs to be addressed.  On the one hand, nearby an extremum of probability distribution, the quadratic behavior dominates in $\phi$.  Then according to Eq.~\ref{eqn_curl_v1}, $\nabla \times \vec{v}_{\rm av}$ is approximately constant, which implies a uniform gyrating direction in this region.  On the other hand, at the special case $k_2 = 8$, our analysis in the last paragraph gives the speculation that the quadratic contribution to the NESS currents is absent (though the quadratic contribution here means that from the probability distribution), and there exist oppositely gyrating regions nearby the extremum at all temperatures.  This paradox can be resolved noting that for the case $k_2=8$, the behavior of $\phi$ near its extremum is quadratic but untilted, as can be observed in the second column of Fig.~\ref{fig_double_well}.  Therefore, according to Eq.~\ref{eqn_curl_v1} the quadratic part in $\phi$ becomes irrelevant in $\nabla \times \vec{v}_{\rm av}$, and the non-quadratic part in $\phi$ serves as the major contributor in the gyrating pattern.

From the above discussions, we find the setting $k_2=8$ gives a special scenario in that the harmonic contribution to the NESS currents vanishes.  From Fig.~\ref{fig_double_well}, one can observe that when $k_2<8$, the gyration nearby the potential minimum is clockwise, and the probability distribution at low temperatures is approximately harmonic, while the semi-major axis is slightly tilted in the clockwise direction.  And when $k_2>8$, the gyration and the tilting of the semi-major axis just exhibit the opposite trend.

\section{Isotropic quartic potential}
\label{sec_iso_quartic}
In the previous section, the double-well potential is locally harmonic nearby each energy minimum.  As a contrast, our second nonharmonic case features an isotropic quartic potential which is entirely nonharmonic.  The potential is defined by
\begin{equation}
  U(\vec{r}) = r^4 = (x_1^2+x_2^2)^2 \,
  \label{eqn_iso_quartic}
\end{equation}
where $r=\sqrt{x_1^2+x_2^2}$ is the distance from the origin.  The potential has a rather flat shape nearby the origin, and it starts to grow up drastically as $r$ further increases (in our case the classification between flat and steep regions depends on the temperature scale).

\begin{figure}[h]
 \centering
 \includegraphics[width=0.5\textwidth]{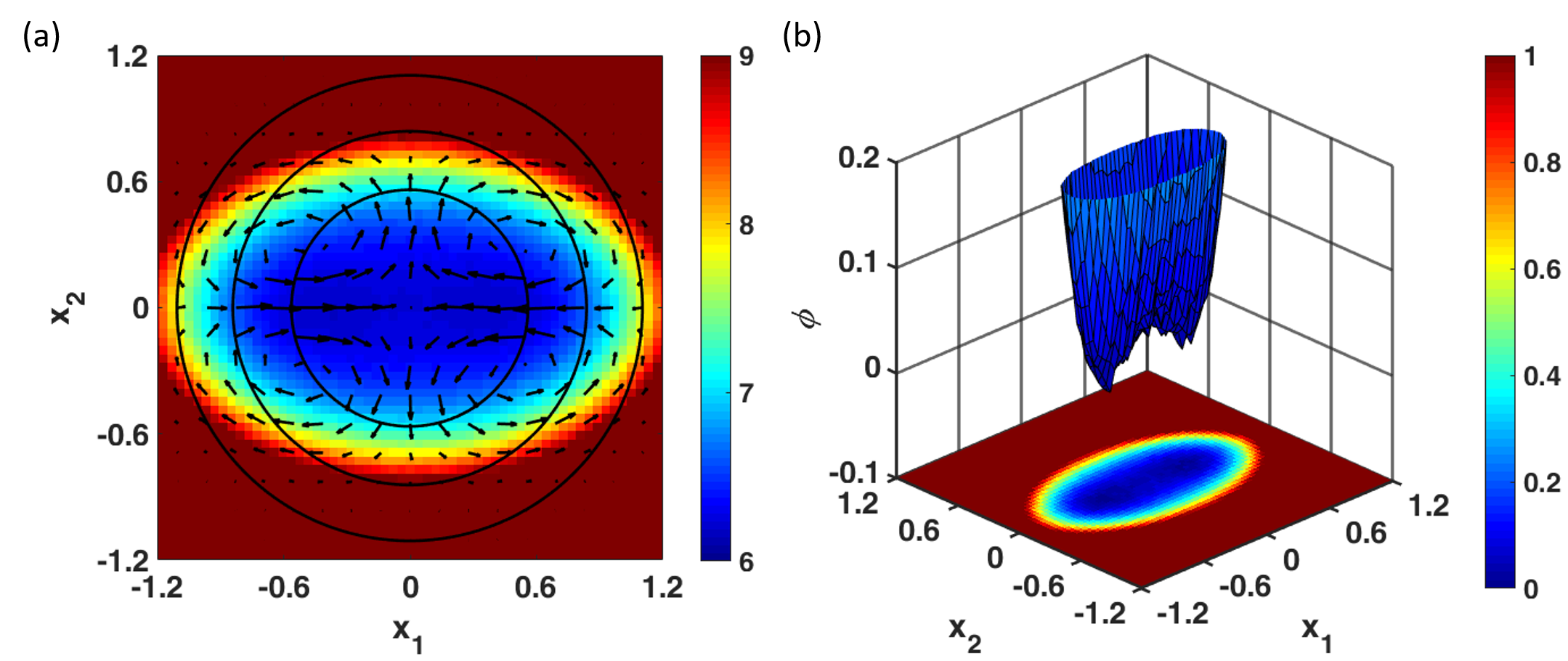}
 \caption{Result for the isotropic quartic potential $U = (x_1^2+x_2^2)^2$.  (a) NESS results with $T_1=0.8$ and $T_2=0.2$ (same notational designation as in Fig.~\ref{fig_harmonic}). (b) The function $\phi = - \log P$ reveals a shallow, untilted double-well structure.  The distribution of $\phi$ is derived using the bin size $x_{\rm bin}=4/99$, while $x_{\rm bin}=4/29$ is adopted in the calculation of $\vec{J}$.  }
 \label{fig_iso_quartic}
\end{figure}

Figure~\ref{fig_iso_quartic} shows the simulation results of the NESS probability and flux density distribution.  Interestingly, the NESS currents exhibit four circulating regions, each residing in its own quadrant, and there is a stationary point of flow within each circulating region.  The origin behaves as a saddle point in terms of NESS dynamics.  Meanwhile, the shape of probability distribution is largely similar to that of the potential profile, and the distribution over the vertical direction is narrower due to the smaller temperature $T_2$.  A more careful examination reveals a shallow double-well structure in $\phi$ (see Fig.~\ref{fig_iso_quartic}(b)), and thus the probability maxima do not occur at the potential minimum (i.e., origin).  Instead, the origin again serves as a saddle point regarding $\phi$ and therefore the probability distribution.

To understand the flowing behavior, we first note that the isotropic quartic potential is even in both $x_1$ and $x_2$, and the probability distribution must possess the same parities.  From Eq.~\ref{eqn_J1} one learns that
\begin{eqnarray}
 J_1(-x_1,x_2) &=& -J_1(x_1,x_2) \nonumber \\
 J_2(-x_1,x_2) &=& J_2(x_1,x_2)  \nonumber \\
 J_1(x_1,-x_2) &=& J_1(x_1,x_2)  \nonumber \\
 J_2(x_1,-x_2) &=& -J_2(x_1,x_2)
 \label{eqn_J_quartic}
\end{eqnarray}
where $J_1$ and $J_2$ are the components of $\vec{J}$.  This property immediately leads to the fact that the NESS currents cannot circulate around the equiprobability contours all the time.  In particular, the NESS currents on the $x_1$ axis cannot have a vertical component, whereas the currents on the $x_2$ axis cannot have a horizontal one.

The gyrating pattern for the NESS currents of this isotropic quartic potential can be further understood using our analysis in Sec.~\ref{sec_circulation}.  First, because $\phi$ is even in both $x_1$ and $x_2$, Eq.~\ref{eqn_curl_v1} tells that $\nabla \times \vec{v}_{\rm av}$ must be odd in $x_1$ and $x_2$.  Moreover, since the NESS profile of $\phi$ has a structure which resembles the potential profile, one can represent it using the crude expression $\phi \approx (a^2 x_1^2 + b^2 x_2^2)^2$ ($a$ and $b$ are constants).  Hence $\nabla \times \vec{v}_{\rm av} \approx 4 a^2 b^2 x_1 x_2$.  Therefore, the direction of circulation, which is capitulated through $\nabla \times \vec{v}_{\rm av}$, is identical in each quadrant, which agrees well with our observation in Fig.~\ref{fig_iso_quartic}.

More precisely, the observed geometry of $\phi$ differs with that of $U$ and our crude approximation of $\phi$ in the fact that it posses a very shallow double-well structure.  We believe that such a mild difference does not result in a big property change in $\nabla \times \vec{v}_{\rm av}$.  In fact, if one adds an harmonic bump upon the existing quartic geometrical shape, Eq.~\ref{eqn_curl_v1} shows that the harmonic term does not change $\nabla \times \vec{v}_{\rm av}$ as long as its shape is not tilted with respect to the $x_1$ axis.  Alternatively, the total entropy analysis predicts that the gyrating direction is dependent on $\partial_1 \partial_2 U = 8x_1 x_2$.  The analysis provides a stronger evidence towards a clockwise gyration in the first and third quadrants and a counterclockwise one in the second and fourth quadrants.

The flow field can also help us get a better understanding in $\phi$ and therefore $P$.  Equation~\ref{eqn_J2} implies that $d\phi  - d\vec{r} \cdot \hat{\bf D}^{-1} \nabla U > 0$, or equivalently,
\begin{equation}
  d\phi > \frac{dx_1 \partial_1 U}{T_1} + \frac{dx_2 \partial_2 U}{T_2}
  \label{eqn_dphi_dU}
\end{equation}
along the NESS flow path $d\vec{r} = (dx_1, dx_2)$.  For the isotropic quartic potential we consider, the NESS flux along the $x_2$ axis directs outward from the origin.  Following this flowing direction along the $x_2$ axis, Eq.~\ref{eqn_dphi_dU} reduces to $d\phi > dU/ T_2$.  On the other hand, if one follows the flowing direction on the $x_1$ axis towards the origin, the criterion gives $d\phi > dU/ T_1$.
Therefore, directing out from the origin, the increase of $\phi$ is sharper than that of $U/T_2$ along the $x_2$ axis and milder than that of $U/T_1$ along the $x_1$ axis.  The change of $\phi$ along the $x_1$ axis can be even negative, which is observed from the shallow double-well structure in $\phi$ (see Fig.~\ref{fig_iso_quartic}(c)).

\section{Total entropy production}
\label{sec_dissipation}
Following a NESS flow path, the rate of entropy change in the heat reservoirs is
\begin{equation}
 \left( \frac{dS_Q}{dt} \right)_{\rm ssf} = \frac{F_1 v_1}{T_1} + \frac{F_2 v_2}{T_2} = - \nabla U\cdot \hat{\bf D}^{-1}\vec{v}_{\rm av}\, ,
\end{equation}
where $F_i \equiv -\partial_i U$.  Meanwhile, the time derivative of the system entropy is
\begin{equation}
 \left( \frac{dS_{\rm sys}}{dt} \right)_{\rm ssf} = - \left( \frac{d\log P}{dt} \right)_{\rm ssf} = \vec{v}_{\rm av} \cdot \nabla \phi \, .
\end{equation}
Thus the total entropy changing rate along the NESS flow path is
\begin{eqnarray}
 \left( \frac{dS_{\rm tot}}{dt} \right)_{\rm ssf} &=& \left( \frac{dS_{\rm sys}}{dt} \right)_{\rm ssf} + \left( \frac{dS_Q}{dt} \right)_{\rm ssf} \nonumber \\
 &=& \vec{v}_{\rm av} \cdot (\nabla \phi - \hat{\bf D}^{-1} \nabla U) \nonumber \\
 &=& \vec{v}_{\rm av} \cdot \hat{\bf D}^{-1} \vec{v}_{\rm av} \geq 0 \, .
 \label{eqn_total_S}
\end{eqnarray}
The last line in Eq.~\ref{eqn_total_S} is derived using Eq.~\ref{eqn_J2}.  Note that the equality in Eq.~\ref{eqn_total_S} holds at stationary points of flow only.  Therefore the total entropy changing rate is positive along the NESS flow path regardless of the type of conservative forces.  Note that Eq.~\ref{eqn_total_S} implies that if one follows the NESS flow path, the total entropy change is just equal to the contribution by the damping force.

If the potential is quadratic, the NESS currents circulate around the equiprobability contours, and the system entropy is a constant of time.  Therefore, the net rate of heat dissipation towards the thermal baths is always positive at the NESS.  On the other hand, regarding non-quadratic potentials, the system entropy becomes oscillatory along a NESS cycle (see Figs.~\ref{fig_double_well} and \ref{fig_iso_quartic} for examples).  Moreover, at times one may encounter a region where the net entropy dissipation into the environment $\Delta S_{Q,{\rm ssf}}$  turns negative.  For example, from Fig.~\ref{fig_iso_quartic} one finds that along the $x_2$ axis, while the NESS current directs away from the origin, the conservative force is just anti-parallel to $\vec{v}_{\rm av}$, and therefore the entropy dissipation into the surrounding is negative.

The above result may look counterintuitive, as during this stage, on average, the system absorbs heat from the low-temperature reservoir.  Meanwhile, following the trajectory along the $x_1$ axis towards the origin, the system (again on average) dissipates heat into the high-temperature reservoir.  The autonomous occurrence of these less intuitive trajectories can be understood by the corresponding changes in the system entropy, which is synonymous with the information possessed by the system, and the thermodynamic law dictates a positive change in total entropy.  Moreover, noting that $\Delta S_{Q,{\rm ssf}}=  \Delta S_{\rm tot, ssf}$ after one full cycle, one deduces that during the rest of the trajectory, behavior of opposite trends must occur.  And after one full cycle, the system indeed absorbs a positive amount of heat from the high-temperature reservoir and dissipate it into the low-temperature reservoir.

\section{Conclusion}
\label{sec_conclusion}
In this work, we apply two simple criteria to examine the gyrating directions.  The first one requires either the input of the NESS probability distribution, or the assumption that its corresponding $\phi$ resembles the potential geometry qualitatively.  While the description of geometric resemblance is itself less well-defined, our second method provides a clearer recipe without {\it a priori} knowledge about the probability distribution.  It is remarkable to note that the second law of thermodynamics plays a substantial role in understanding the gyrating dynamics\cite{RC_gyrator_2017}.

For the harmonic potential, the system possesses the remarkable feature that the NESS currents gyrate about the equiprobability contours.  As for the double-well potential, oppositely gyrating regions exist nearby each potential minimum, and the gyrating direction where the potential minima reside depends on the choice of $k_2$.  At the critical value $k_2=8$, the equipotential contours nearby the potential minima become circular, and the harmonic contribution to gyrating pattern vanishes completely.  Finally, for the isotropic quartic potential, which is completely nonharmonic, the result reveals four circulating regions, and the flowing pattern can be comprehended using simple arguments in parity.

%The resulting $\phi$ gives a very shallow double-well shape.  We believe that these locally harmonic structures are irrelevant in the overall gyrating pattern due to their untilted shape.

Our observations based on this work and other trial potentials lead us to speculate that, unlike harmonic potentials, the NESS currents do not faithfully follow the tangent of equiprobability contours for general nonharmonic potential cases.  Nevertheless, in our nonharmonic results one can notice areas where this circulating-about-probability-contour feature is seemingly present (see Figs.~\ref{fig_double_well} and \ref{fig_iso_quartic}, away from the interface between oppositely gyrating regions).  Does the nonharmonic potential create extra vortices that locally break the gyrating feature $\vec{v}_{\rm av}\cdot \nabla \phi =0$?  This na\"ive speculation is left to be further examined in future works.
With a comprehensive understanding in the interplay between the nonharmonic potentials and their resulting gyrating patterns, one can anticipate more promising ideas in the advances of Brownian engines.
%Note that our simple criteria cannot predict the positions of stationary pointsof flow.  While some perturbative methods\cite{Kwon_PRE2011, Grier_PRE2015} have been developed on the studies of general NESS dynamics beyond harmonic approximations, it remains important to build a simple approach that can study the NESS characteristics beyond local approximations.

%%%%%%%%%%%%%%%%

\begin{acknowledgments}
%The authors wish to thank Ping Ao for stimulating discussions.
This work has been supported by Ministry of Science and Technology in Taiwan under grant MOST 107-2112-M-008-014.  C.-L.L. acknowledges the support from NCTS thematic group Complex Systems.
\end{acknowledgments}


\begin{thebibliography}{99}
\bibitem{footnote}
In this work we use the term ``stationary point'' to designate the point where the gradient of some scalar function is zero, while we use the term ``stationary point of flow'' to designate the point where $\vec{v}_{\rm av}=\vec{0}$.
\end{thebibliography}
\end{document}